\documentclass[twocolumn,prb,aps,epsf,showpacs]{revtex4-1}

\usepackage{graphicx}
\usepackage{dcolumn}
\usepackage{bm}

\begin{document}

\title{Energy Gap and Spin Polarization in the 5/2 Fractional Quantum Hall Effect }

\author{S. Das Sarma$^{1}$, G. Gervais$^{2}$, and Xiaoqing Zhou$^2$}

\affiliation{$^{1}$ Condensed Matter Theory Center, Department of
Physics, University of Maryland, College Park, MD 20742 USA}

\affiliation{$^{2}$ Department of Physics, McGill University,
Montreal, H3A 2T8, CANADA}

\date{\today }

\begin{abstract}

We consider the issue of the appropriate underlying wavefunction describing the enigmatic 5/2 fractional quantum Hall effect (FQHE), the only even denominator FQHE unambiguously observed in a single layer two dimensional (2D) electron system. Using experimental transport data and theoretical analysis, we argue that the possibility of the experimental 5/2 FQH state being not fully spin-polarized cannot be ruled out. We also establish that the parallel field-induced destruction of the 5/2 FQHE arises primarily from the enhancement of effective disorder by the parallel field with the Zeeman energy playing an important quantitative role.

\end{abstract}
 \maketitle

\section{Introduction}

The observation of the 5/2 FQHE, with two completely filled Landau levels (spin up and down) belonging to the lowest orbital (N=0) and the half filling of the second (N=1) orbital Landau level (SLL), has remained an intriguing enigma 23 years after its discovery\cite{Willet87}. The 5/2 FQHE and its conjugate state (at 7/2 = 6 - 5/2) are the only known even denominator FQHE unambiguously observed in a single-layer 2D system. All other $\sim 80$ observed FQH states obey the odd-denominator rule,  consistent with the Pauli principle,  as originally described by Laughlin in his seminal theory of the  1/3 FQHE\cite{Laughlin83}. \\


In addition to the intriguing even-denominator nature of the 5/2 FQHE, the other puzzling aspect of the 5/2 state is its existence in the SLL where very few (less than 10) FQH states have been observed,  in sharp contrast to the $N = 0$ lowest Landau level (LLL) where $\sim$70 FQH states have been established\cite{Pan07PRB}. The even denominator filling fractions $\nu = 1/2$ and $\nu =3/2$ in the LLL do not manifest any incompressible FQHE, and are instead compressible Fermi liquid states.  Another characteristic feature of the 5/2 FQHE is its very fragile nature with a small measured activation gap $\sim$0.1 - 0.5~K, even in the world's highest mobility (up to $3.5 \times 10^7 cm^2/V\cdot s$) samples. The 5/2 FQHE is thus observed only at very low temperature ($\lesssim 100$ mK) and in samples with very high mobility ($\mu > 10^7 cm^2 / V\cdot s$). These constraints (very low temperature and very high mobility) have limited the experimental investigation of the 5/2 FQHE and almost twenty five years  after its discovery, the precise nature of the observed 5/2 FQHE and its underlying theoretical description are still a subject of active debate.\\

There is, however, an almost consensus $theoretical$ candidate for the 5/2 FQHE, the Moore-Read Pfaffian (Pf) wavefunction\cite{MR90}, which is a chiral spinless $p+ip$ paired BCS state for the composite fermions at 5/2 filling. The low-lying quasiparticle excitations of this state are known to be  Ising anyons obeying the $(SU2)_2$ conformal field theory universality class\cite{Nayak08RMP}. Being a paired state of fermions, the Pf can have an even denominator since it is not constrained by the Pauli principle. There has been a great deal of recent theoretical and experimental interest in the 5/2 FQHE following the concrete suggestion of the construction of a topological 5/2 qubit\cite{Sankar05QC} using the braiding of the non-Abelian quasiparticles of the Pf state, provided  the observed state is indeed the proposed theoretical non-Abelian  Pf state. (All our discussions in this work with respect to spin-polarization of the 5/2 state and the appropriate theoretical description apply equally well to the so-called anti-Pf non-Abelian wavefunction also, since the anti-Pf state has the same bulk properties as the Pf state, and is also thought to be completely spin-polarized-- we refer the reader to Refs.\cite{Nayak08RMP,Radu08} for a discussion of the possible distinction between the edge properties of the Pf and the anti-Pf theoretical candidate states.) \\

Recent experimental work \cite{Radu08} performed on the edge, in a narrow constriction, has shown the inter-edge  tunneling to be consistent with a 5/2 FQHE emanating  from a non-abelian spin-polarized ground state. Also exciting is the recent observation of interference patterns \cite{Willet09} which alternate between those of quasiparticles with $e^*=e/4$ and $e^*=e/2$. This interference pattern has been interpreted as being equivalent to the observation of non-Abelian statistics although there exists no satisfactory theory explaining the observations. While these edge experiments are encouraging, there remain complications associated with the edge, such as edge reconstruction and coupling between bulk quasiparticles and the edge, which in the end can make their interpretation difficult. The question that arises naturally is whether or not there is any evidence at all for a Pf state at 5/2 filling in the bulk of a high-mobility two-dimensional electron gas (2DEG).
Experiments aimed at direct shot-noise-based measurements of the quasiparticle charge in the 5/2 FQHE state have given ambiguous and non-universal results\cite{Dolev2008,Dolev2010} varying in a complicated manner with temperature and tunneling strength, which are inconsistent with the bulk 5/2 state being a non-Abelian Pf state. \\

There is however compelling theoretical evidence\cite{Storni10}, based on numerical studies of small systems containing a few (8 - 20) electrons, that the actual experimental state in the presence of realistic inter-particle Coulomb interaction is indeed adiabatically connected to the Pf state, however there is at the moment no experimental evidence showing that the bulk is indeed described by the Pf wavefunction. We argue below, based on all
energy gap measurements of the 5/2 FQHE available in the literature, that the experimental evidence 
for the 5/2 state (in the bulk) tends more towards a  spin unpolarized rather than a spin-polarized state, in contradiction with the prediction from a Pf state.  \\

The specific physical question which is the primary topic of the current work relates to the nature of the spin polarization of the 5/2 FQH state. In particular, the Pf wavefunction is fully spin-polarized, and direct numerical work\cite{Morf98, Feiguin09PRB} indicates that the exact small-system state has lower ground state energy in the fully spin-polarized case than in the unpolarized case, although the energy difference between the two is small. Since the spin-polarized Pf wavefunction also has relatively good overlap, albeit not excellent, with the exact numerical small system wavefunction, the theoretical consensus has almost universally been that the experimentally observed 5/2 FQH state is {\it i)} fully spin-polarized,  and {\it ii)} described by the Pf wavefunction (or more precisely, belongs to the same non-Abelian universality class as the Pf wave function and is adiabatically connected to the Pf).\\

In sharp contrast to the theoretical consensus based on small-system numerical work (which can, in principle, be questioned since the experimental system has $\sim10^9$ electrons whereas  numerical work is typically based on $\sim 10$ electrons), there has been absolutely no direct experimental evidence supporting the claim of full spin-polarization of the 5/2 FQH state. In fact, existing (mostly circumstantial) experimental evidence hints towards a spin unpolarized 5/2 FQHE. First, the application of a parallel magnetic field ({\it i.e.} along the 2D plane without affecting the Landau quantization), which presumably varies the Zeeman energy, is found to rapidly destroy the 5/2 FQHE even for a relatively modest field. Second, the 5/2 FQHE is found to exist  at rather low magnetic field (down to 2.5~T); in fact, the measured 5/2 FQHE activation gap ($\sim$0.25~K) at 2.5~T\cite{Dean08}  is {\it larger}  than that at 10.5~T($\sim0.1$~K) \cite{Zhang10}, which seems to imply that the state is spin unpolarized since it is weakening with increasing magnetic field (whereas the Coulomb energy by itself increases with increasing magnetic field). Motivated both by the fundamental importance of the question and the stark dichotomy between the theoretical consensus and the experimental lack of evidence, we critically revisit the issue of 5/2 spin-polarization using the existing experimental data as well as some new transport data of our own. Our analysis rather supports the conclusion that the experimentally observed 5/2 FQH state is  more likely to be  spin-unpolarized than spin-polarized. Our conclusion, along with recent optical and light scattering measurements \cite{Stern10, Pinczuk07}, throws the subject of the nature of  the 5/2 FQHE into serious jeopardy, necessitating a rethinking of the possible spin unpolarized 5/2 FQHE candidate states. In addition, it is known that at higher temperatures the 5/2 incompressible FQHE goes into a compressible Fermi liquid phase which appears to be spin-unpolarized\cite{Willet02}. The same absence of spin polarization is also found for the compressible $\nu =1/2$ LLL state\cite{Tracy07}.\\


\section{Analysis}

Our analysis of experimental data follows two related but distinct tracks. First, we consider the existing activation gap measurements as a function of the (perpendicular) magnetic field $B_{\bot}$ to verify whether the measured gap $\Delta(B_{\bot})$ is more consistent with the spin-polarized or unpolarized 5/2 FQHE. Second, we consider the parallel field ($B_{\parallel}$) induced suppression $\Delta(B_{\parallel})$ of the 5/2 FQHE to verify its consistency (or not) with a spin unpolarized 5/2 ground state. Making the standard assumption that the activation gap $\Delta$ essentially measures an excitation gap, we can write down the following simple general formula for the measured gap, $\Delta(B_{\bot}, B_{\parallel})$,  as a function of the perpendicular $B_{\bot}$ and the parallel magnetic field $B_{\parallel}$, 
\begin{equation}
\Delta = a_1 f_1(B_{\bot}, B_{\parallel})\sqrt{B_{\bot}} - a_2 \gamma(B_{\parallel}) - a_3 g \mu_0\sqrt{B_{\bot}^2+B^2_{\parallel}}.
\label{eq:gap}
\end{equation}  

The three terms in Eq.~\ref{eq:gap} correspond respectively to the increasing incompressible gap with the perpendicular field due to the increasing Coulomb energy as $\l_{B}^{-1}$ (where $l_{B}=\sqrt{\hbar c/eB_{\bot}}$ is the magnetic length) modified by the finite width effect (parameterized by the $f_1$ function) which suppresses\cite{Petersen08B} the Coulomb energy, the disorder effect suppressing the gap which may increase with the parallel field, and finally the Zeeman energy term which is proportional to the total magnetic field $B_{tot}\equiv \sqrt{B^2_{\bot}+B^2_{\parallel}}$. The finite width correction parameterized by the function $f_{1}(B_{\bot},B_{\parallel})$ in the first term is usually only a small correction ($\sim10-20\%$). The coefficient $a_1=C_{5/2}e^2/(\epsilon \sqrt{\hbar c/e})$, where $\epsilon$ is the GaAs background lattice dielectric constant, and theoretically $C_{5/2}=0.025$ for the 5/2 FQHE according to exact numerical calculations. The disorder effect, parameterized by the second term is known\cite{Morf04} to be extremely important quantitatively for the 5/2 FQHE. In the absence of any applied parallel field, $\gamma(B_{\parallel})=1$ and $a_2 = \Gamma_0$, where $\Gamma_0$ is the gap suppression due to the disorder broadening of the sample. It is customary to take $\Gamma_0$ as the single-particle broadening defined by the Dingle temperature, $\Gamma_0 = T_D$, as measured by the low field SdH oscillations. We adopt this idea in our analysis. Finally, in the third term $g\mu_0$ denotes the Zeeman splitting strength (with $g$ the Lande $g$-factor and $\mu_0$ the Bohr magneton) with the coefficient $a_3$ controlling the nature of spin polarization of the system. For a spin-polarized ground state, $a_3=0$ unless the relevant low-lying excitations are spin-reversed in which case $a_3=-1$. For a spin unpolarized ground state, where the Zeeman energy should suppress FQHE, we choose $a_3=+1$. In principle, $g$ can be used as an unknown parameter for a partially spin polarized state, but we take $g=0.44$ (unless otherwise stated) to be consistent with the known value of the $g$-factor in GaAs. Below we will consider two distinct cases separately: $B_{\parallel}=0$, $a_{2}=T_{D}$,  $\gamma=1$ (Fig.1),  and  $B_{\parallel}\neq 0$ (Fig.2). \\



Before presenting a detailed comparison between Eq.~\ref{eq:gap} and the experimentally measured activation gap, we point out that there is a qualitative difference between  spin-polarized ($a_3=0$) and unpolarized ($a_3=1$) situations according to Eq.~\ref{eq:gap}. In particular, assuming no parallel field, {\it i.e.} $B_{\parallel}=0$ and  $\gamma(B_{\parallel})$=1,  and a $B_{\bot}$-independent disorder broadening (a standard approximation in most theoretical FQHE analysis), Eq.\ref{eq:gap} implies a maximum  in $\Delta$ as a function of $B_{\bot}$ at a value $B_{\bot}=(\frac{a_{1}}{2g \mu_{0}})^2\simeq 4.5T$ for the spin unpolarized case due to  the increasing Coulomb energy  $\sim\sqrt{B_{\bot}}$ that opposes  the decreasing Zeeman term ($\sim B_{tot}$). There is no such maximum as a function of $B_{\bot}$ for the spin polarized case ($a_3$=0). This maximum as a function of $B_{\bot}$ translates directly into a maximum in the activation energy as a function of the sample density ($n$) since the filling factor  is fixed ($\nu=5/2$). Then, a clear prediction of the theory is that if the 5/2 state is spin unpolarized, then the corresponding activation gap would show a maximum as a function of electron density at a density of $n\simeq 3.5 \times 10^{11}cm^{-2}$. On the other hand, a spin polarized 5/2 FQHE will manifest an activation gap increasing monotonically with $B_{\bot}$. Although the actual experimental situation shown in Fig.1 for $B_{\parallel}=0$ is considerably compromised by disorder effects which depend on density, we believe that the whole collection of the existing experimental data on the 5/2 FQHE is more consistent with there being a maximum in the measured activation gap around a density of $3\times 10^{11}cm^{-2}$. In particular, the measured activation gap at $B_{\bot}\sim 10.5~T$ by Zhang {\it et al.} \cite{Zhang10} for a sample with electronic density $n=6.2 \times 10^{11} cm^{-2}$ is a factor of two smaller than that at $B_{\bot}\sim 2.5~T$\cite{Dean08} at a density $n=1.6 \times 10^{11}cm^{-2}$,  in spite of the two samples having very comparable mobilities. This trend is more consistent with the FQHE being spin-unpolarized than spin-polarized. Conversely, if the 5/2 FQHE is considered to be spin-polarized, one would have expected a very large measured gap at 10.5~T. In addition, the early measurements of Pan {\it et al.}\cite{Pan01}, although not decisive, seem to clearly indicate a shallow maximum, as implied by Eq.~\ref{eq:gap} with $a_3 = + 1$ ({\it i.e.} spin unpolarized 5/2 FQHE), in the activation gap data\cite{Eisenstein_PC2}.\\

\begin{figure}
\includegraphics[width=1.0\linewidth]{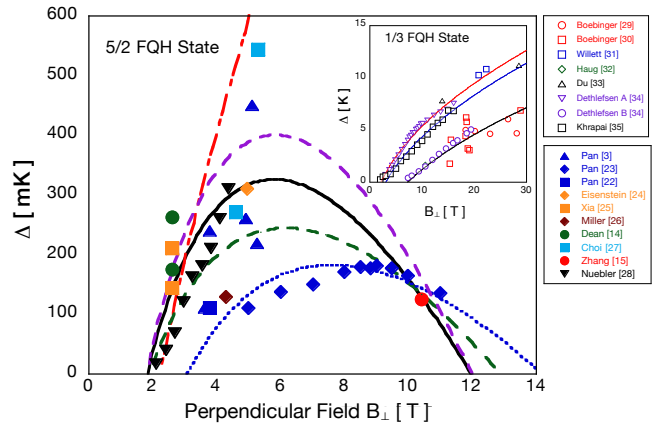}
\caption{The activation energy gap the 5/2 FQH state  measured by various groups (filled symbols) is shown versus the perpendicular field \cite{Pan07PRB,Pan99B,Pan01,Eisenstein02, Xia10,Miller07,Dean08, Choi08, Zhang10, Nuebler10}. Assuming $f_1=1$ ({\it i.e.} not including the finite width correction), the trend of these data points can be fit to Eq.~\ref{eq:gap} with the following parameters: $a_1=1421$, $T_D=1387$, $a_3=1$ and $g=0.44$(solid black);  $a_1=1000$, $T_D=1000$, $a_3=1$ and $g=0.30$(dashed green); $a_1=1000$, $T_D=1200$, $a_3=1$ and $g=0.27$(dotted blue); $a_1=800$, $T_D=1200$ and $a_3=0$(dot-dashed red). Including the finite width correction for a typical 30 nm QW, new set of fit parameters $a_1=1600$, $T_D=1600$ and $g=0.44$ (dashed purple) are needed for a better fit, but the qualitative curve shape remains the same. The inset shows the energy gaps for the spin-polarized 1/3 FQH state determined using transport measurements \cite{Boebinger85,Boebinger87,Willet88,Haug87,Du93,Dethlefsen06} and direct measurement of the chemical jump\cite{Khrapai08}. These data points are fit to Eq.~\ref{eq:gap} with the following parameters: $a_1=3.14$, $T_D=4.57$ and $a_3=0$(solid red); $a_1=3.05$, $T_D=5.30$ and $a_3=0$(solid blue); $a_1=2.56$, $T_D=6.88$ and $a_3=0$(solid black).}
\label{fig:1/3}
\end{figure}

In Fig.\ref{fig:1/3} we compare all the existing 5/2 FQHE activation data with Eq.~\ref{eq:gap}, assuming a few distinct parameter values (since $T_D$ is not always reported for all the samples used in FQHE activation measurements). Although no compelling decisive conclusion is possible, we believe that the comparison gives a  preference to the 5/2 FQH state being spin unpolarized ({\it i.e.} $a_3=1$) rather than spin polarized ($a_3=0$). In particular, the expected monotonic increase of the gap ($\sim\sqrt{B_{\bot}}$) with increasing magnetic field for the spin-polarized situation is simply not apparent in the data. Conversely, all of the data taken together (using samples from different groups) is more consistent with the gap first increasing with $B_{\bot}$ and then decreasing at large $B_{\bot}$ with a shallow maximum around $B\approx 5.5~T$, which is consistent with the 5/2 state being spin unpolarized . We emphasize that we can only discuss the general trend of the data as being more consistent with the 5/2 state being spin-unpolarized since $\Delta$ seems to be typically smaller at larger $B_{\bot}$ than at smaller $B_{\bot}$. Of the two sets of existing data in a density-tunable sample, the data from Pan {\it et al.}\cite{Pan01} clearly manifest a shallow maximum indicating the lack of spin polarization whereas the data from Nuebler {\it et al.}\cite{Nuebler10} exists only over a small range of low $B_{\bot}$ values to draw any decisive conclusion. Last, to emphasize the unexpected behaviour of the 5/2 gap with the perpendicular field, we show as an inset of  Fig.~\ref{fig:1/3} the magnetic field dependence of the 1/3 energy gap measured by various groups. As expected for a spin-polarized FQH state, the 1/3 energy gap increases monotonically with $B_{\bot}$. \\

Now we turn to the consideration of the measured activation gap in an applied parallel field ($B_{\parallel}\neq 0)$. Everything else being equal, $\Delta$ for a spin-polarized situation should either increase with or not exhibit any dependence on the parallel magnetic field ($B_{\parallel}$). On the other hand $\Delta$ for the spin-unpolarized situation should decrease monotonically with increasing $B_{\parallel}$. This is precisely the experimental observation for all 5/2 FQHE studied in the literature. The truly puzzling phenomenon, however, is that the measured activation gap $\Delta$ decreases with increasing $B_{\parallel}$, also for other FQH states, {\it e.g.} $\nu=1/3$, $2/5$, where the ground state (e.g. $\nu=1/3$) is most definitely spin-polarized. This leads to the conundrum that the parallel field induced destruction of the 5/2 FQHE  {\it cannot} be naively attributed entirely to its spin-unpolarized nature since fully spin-polarized FQH states can also be suppressed by a large applied $B_{\parallel}$. We propose that this additional mechanism is the increasing of effective disorder broadening in the parallel field due to the magneto-orbital coupling\cite{DasSarma2000}, which is particularly strong in a wide sample with well width $d \gg l_{B}$. Experimentally, a crude estimate of $B_{\parallel}$ induced enhancement of effective disorder is given by the measurement of $\rho(B_{\parallel})$ in zero perpendicular field ($B_{\bot}=0$). Using the measured $\rho(B_{\parallel})$, we can crudely approximate the function $\gamma(B_{\parallel})$ in Eq.~\ref{eq:gap}, which is taken to be unity ($\gamma=1$) so far in our analysis, to be

\begin{equation}
\gamma(B_{\parallel}) = \frac{\rho(B_{\parallel})}{\rho(0)},
\label{eq:gap3}
\end{equation}

\noindent where we are making the reasonable assumption that the effective disorder scales with the measured 2D resistivity in the parallel field. Experimental measurements\cite{Zhou10} show that $\gamma(B_{\parallel})\gg1$ as defined by Eq.~\ref{eq:gap3}, at least for wide samples at high magnetic field.

\begin{figure}
\includegraphics[width=1.0\linewidth]{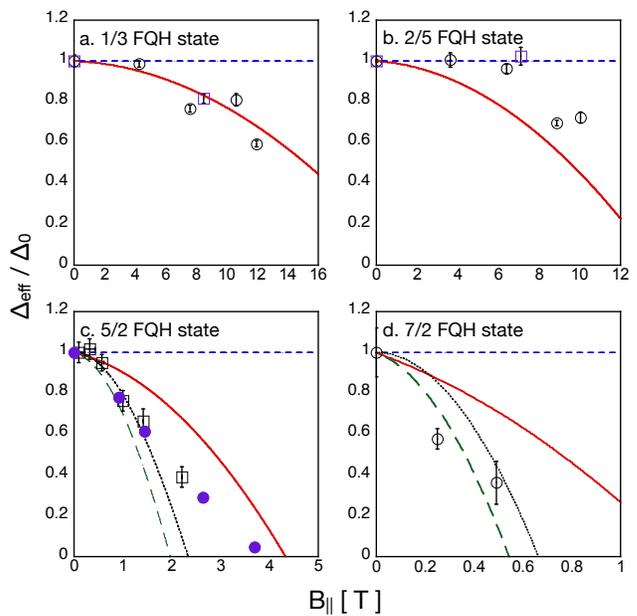}
\caption{Suppression of the activation energy gap in a tilted field for the a) 1/3 \cite{Piot08}; b) 2/5  \cite{Piot08}; c) 5/2 \cite{Dean08,Xia10} and d) 7/2  \cite{Dean08} FQH states, all measured in a 40 nm sample. The data for the 1/3 and 2/5 FQH states \cite{Piot08} are shown for different cooldown with square and circle symbols. For the 5/2 state, data from Ref. \cite{Xia10} measured in the same wafer were added and are shown as filled circles.  The gap suppression calculated from  Eq.~\ref{eq:gap} is shown by the lines, using the following parameters: $g=0$, $T_D=230~mK$ (solid red); $g=0$, $T_D=0~mK$ (dashed blue); $g=0.44$, $T_D=230~mK$ (dashed green); $g=0.44$, $T_D=0~mK$ (dotted black). The disorder parameter, $\gamma(B_{\parallel})$, was extracted from a second order polynomial fit to the low-field (0-6T) parallel field magnetoresistance \cite{Zhou10}.}
\label{fig:2}
\end{figure}

Using the measured $\gamma(B_{\parallel})$ and the Dingle temperature $T_{D}$,  as  well as the measured gaps for $\nu = 5/2$, 7/2, 1/3, and 2/5 all from the same 40 nm wide quantum well \cite{Dean08,Piot08,Xia10}, we show in Fig.\ref{fig:2}  the comparison between the theoretically expected behaviour of the activation gap $\Delta(B_{\parallel})$ and the experimentally measured gap, as a function of $B_{\parallel}$. It is apparent that a large part of the strong decrease of $\Delta$ with increasing $B_{\parallel}$ can be attributed to the increasing effective disorder, as reflected by the increasing $\rho(B_{\parallel})$ in the 2D system. The fact that the gap suppression as a function of $B_{\parallel}$ happens not only for the enigmatic 5/2 (and the 7/2) state, but also for the manifestly spin-polarized 1/3 (and 2/5) state lends credence to our idea of an effective disorder induced gap suppression in an applied parallel magnetic field.  We note that there is one well-known exception to the behavior shown in Fig.\ref{fig:2}: the 7/3 gap in a 40 nm sample increases with increasing $B_{\parallel}$ as observed by two groups\cite{Dean08, Xia10}, and we have no explanation for this observation except to note that somehow spin-reversed excitations must play an important role for the 7/3 FQHE in this sample.\\

We now comment on the important question of the implication of Fig.\ref{fig:2} for the issue of spin-polarization of the 5/2 FQHE. As can be seen in Fig.\ref{fig:2}, taking $g=0$, {\it i.e.} assuming the system to be spin-polarized (red line), the effective disorder model by itself provides a  reasonable explanation for
the suppression of the 1/3 and 2/5 gaps by the parallel field. On the other hand, taking $g=0$ does not fully account for the suppression of the 5/2 gap; the effective disorder  with increasing $\gamma(B_{\parallel})$ only accounts for a fraction of the decrease of $\Delta(B_{\parallel})$. This difference could arise from a $B_{\parallel}$-induced Zeeman contribution further weakening the gap. In Fig.\ref{fig:2}, we show theoretical results including the Zeeman term (dashed green), which tends to give better agreement between the theoretical plots and the experimental suppression of $\Delta(B_{\parallel})$. We therefore conclude that for the 5/2 (and 7/2) FQHE, both effective disorder and Zeeman energy contribute to the observed gap suppression for $B_{\parallel}\neq 0$ whereas for the established spin-polarized FQHE ($\nu=1/3,2/5$), the suppression arises most likely from the enhanced disorder in a parallel field.\\

\section{Discussion}

We now comment on the importance of  our theoretical modelling for the determination of the spin polarization via energy gap measurements. From the data in Fig.\ref{fig:1/3}, it is clear that new data for the 5/2 FQH energy gap are needed above $\sim$5.5 T in order to conclude on the spin polarization from the dependence of the energy gap on the magnetic field and density. This is of paramount importance  because Eq.\ref{eq:gap} makes the simple prediction of a positive (negative) slope for the energy gap $\partial \Delta /\partial B_{\bot}$ in the case of a spin polarized (unpolarized) state. This experiment can in principle be performed in  an ultra-high mobility tunable device where the electron density $n$  is swept continuously in the range between $\sim$6 to $\sim$10T, and the slope of the energy gap $\partial \Delta /\partial n$ determined. Recent experiments using optical spectroscopy \cite{Stern10,Pinczuk07} have shown evidence for the 5/2 state to be spin unpolarized, up to the lowest temperature
probed in these experiments. This lack of spin polarization was also found  for other FQH states
in the SLL. If these measurements are correct, then our model predicts a maximum to occur for the energy gap
of {\it any} spin unpolarized FQH states. Therefore, a determination of the energy gap over a wide range of electron densities and for several FQH states in the SLL is at the moment of the utmost importance in order to determine whether or not  the 5/2 state is spin polarized as theory predicts it to be.\\

What we are emphasizing in this work is the lack of the clear monotonic increase of the measured energy gap with the magnetic field (or carrier density) in existing experiments as would be required for the 5/2 experimental FQHE to be spin-polarized.  In fact, we believe that there is some evidence for the experimentally measured 5/2 FQH gap to be manifesting a shallow maxima as a function of magnetic field, which would be consistent with a spin-unpolarized state. Only more experiments carried out on a single sample with a tunable carrier density can satisfactorily resolve this question since comparison among different samples with different densities and disorder may not be very meaningful.  We do add a caveat here regarding the effect of the finite width of the sample which would affect the first term in Eq.\ref{eq:gap}  ({\it i.e.} the function $f_1$) and which we have neglected so far in our consideration.  If the effective 2D width of the sample is much larger than the magnetic length $l_{B}$, which would happen for either very thick 2D samples or for very low magnetic fields, then the energy gap would no longer increase monotonically with the applied field in a square-root fashion, instead manifesting an almost saturation behavior with increasing field \cite{Zhang86,PetersonPRB}. This finite width effect would have to be taken into account in the theoretical modelling if and when experimental results measuring the 5/2 FQHE energy gap in a single density-tunable system becomes available.\\

We mention finally that even if the 5/2 FQHE turns out eventually not to be completely spin-polarized, this does not necessarily imply the inapplicability of the Pf wave function as the underlying description.  It is possible to construct a non-Abelian Pf wave function for a partially spin-polarized state.  In the unlikely scenario that the experimental 5/2 FQHE turns out to be completely spin-unpolarized, one would have to seriously consider Abelian candidates, such as the Halperin 331 strong-pairing wave function which is known to be an excellent description\cite{He1993} for the two-component even-denominator FQHE observed frequently at half-filled Landau levels in bilayer systems.


\section{Conclusion}

To conclude, we have critically analyzed the activation gap measurements for the 5/2 (and 7/2) FQHE, 
and found that the large body of experimental gap measurements is more consistent with the 5/2 FQHE being spin-unpolarized than spin-polarized. While the question of the 5/2 spin-polarization cannot be settled without a direct measurement of the spin-polarization itself,  our critical analysis makes such measurements all the more urgent. A clear prediction of the present analysis is that the 5/2 activation gap for a spin unpolarized FQHE should at first increase with sample density and then decrease, manifesting a shallow maximum at  some intermediate density, whereas the corresponding spin-polarized gap will manifest a monotonic increase with sample density (until the density is high enough so that finite sample width effects come into play in affecting the inter-particle Coulomb interaction). We also predict that the observed suppression of the 5/2 activation gap in an applied parallel magnetic field should be weaker in narrower 2D samples because of weaker magneto-orbital coupling. Finally, given the great importance of the tentative conclusion about the nature of the spin-polarization of the 5/2 FQHE with fundamental consequence for its non-Abelian nature, we hope that new experiments will be undertaken to resolve the nature of spin-polarization of the 5/2 FQHE state. 

\begin{acknowledgments}
This work has been supported by NSERC (Canada), CIFAR, FQRNT (Qu\'ebec) and Microsoft Station-Q.

\end{acknowledgments}

\end{document}